\documentclass{PoS}

\usepackage{latexsym} 	
\usepackage{amsmath}  	
\usepackage{amsbsy}
\usepackage{epsfig}     
\usepackage{cite}

\newcommand{\Tc}{\mathrm{T}_c}
\newcommand{\md}{\mathrm{d}}
\newcommand{\rec}{\mathrm{rec}}

\newcommand{\RGsub}{G^{\rm sub}/G^{\rm sub}_{\rm rec}}
\newcommand{\RGdiff}{G^{\rm diff}/G^{\rm diff}_{\rm rec}}

\newcommand{\cO}{{\cal O}}

\newcommand{\be}{\begin{equation}}
\newcommand{\ee}{\end{equation}}
\newcommand{\bea}{\begin{eqnarray}}
\newcommand{\eea}{\end{eqnarray}}
\newcommand{\bean}{\begin{eqnarray*}}
\newcommand{\eean}{\end{eqnarray*}}

\newcommand{\Jpsi}{J/\psi}

\title{
\vspace*{-2.6cm}
\begin{flushright}\texttt{\footnotesize
BI-TP 2010/39\\}
\end{flushright}
\vfill
Charmonium correlation and spectral functions at finite temperature }

\ShortTitle{charmonium correlation and spectral functions at finite temperature }

\author{\speaker{H.-T. Ding}\footnote{Present address: Physics Department, Brookhaven National Laboratory, Upton, NY 11973, USA.}, A. Francis, O. Kaczmarek and H. Satz
\\
Fakult\"{a}t f\"{u}r Physik,
 Universit\"{a}t Bielefeld, D-33615 Bielefeld, Germany\\
     E-mail: \email{hengtong.ding,afrancis,okacz,satz@physik.uni-bielefeld.de}
     }

\author{F. Karsch\\ 
Physics Department, Brookhaven National Laboratory, Upton, NY 11973, USA\\
 Fakult\"{a}t f\"{u}r Physik,
 Universit\"{a}t Bielefeld, D-33615 Bielefeld, Germany\\
        E-mail: \email{karsch@physik.uni-bielefeld.de}}

\author{W. S\"oldner \\
Frankfurt Institute for Advanced Studies, J. W. Goethe Universit\"at Frankfurt, D-60438
Frankfurt am Main, Germany \\
GSI Helmholtzzentrum f\"ur Schwerionenforschung, D-64291 Darmstadt, Germany\\
        E-mail: \email{w.soeldner@gsi.de}}

\abstract{We study the properties of charmonium states at finite temperature in quenched QCD on isotropic lattices. We measured charmonium correlators  using non-perturbatively $\cO(a)$ improved clover fermions on fine ($a=0.01$ fm) lattices with a relatively large size of $128^{3}\times 96$, $128^3\times48$, $128^3\times32$ and $128^3\times24$ 
at $0.73~T_c$, $1.46~T_c$, $2.20~T_c$ and $2.93~T_c$, respectively. Our analysis suggests that $\Jpsi$ is melted already at $1.46~T_c$ and $\eta_c$ starts to dissolve at $1.46~T_c$ and does not exist at higher temperatures. We also identify the heavy quark transport contribution at the spectral function level for the first time. 
}

\FullConference{The XXVIII International Symposium on Lattice Filed Theory\\
		 June 14-19, 2010\\
		 Villasimius, Sardinia Italy}

\usepackage{epsfig}
\usepackage{graphicx}
\usepackage{dcolumn}
\usepackage{bm}

\begin{document}

\maketitle


Considerable interests on the charmonium states in the medium~\cite{Rapp08} have been triggered ever since the proposal of considering the dissolution of the charmonium states as a signal of the formation of QGP~\cite{Satz86}. However, the understanding of the fate of the charmonium states in the medium still remains very difficult. The potential model approach at finite temperature is still under scrutiny since the potential used in the Schr\"odinger equation is not rigorously defined~\cite{Mocsy09}. Remarkable progress has been made recently in the heavy quark effective theory at finite temperature~\cite{Brambilla10}, where the potential can be rigorously defined. However, it requires scales to be hierarchically  ordered. More recently, a new approach based on the path integral formalism has been proposed~\cite{Beraudo10}. It needs further research since the effects of the medium on the heavy quark in this approach is modeled with only Coulomb interaction. First principle calculations in lattice QCD are thus crucially needed to determine the properties of the charmonium states in the hot medium.


  With the lattice QCD approach~\cite{Bazavov09}, the properties of the charmonium, which can be directly seen from the spectral function, are contained in the Euclidean time correlation functions. The extraction of spectral functions from correlators is rather difficult due to the limited number of points in temporal direction required to perform an analytic continuation from imaginary to real time. In this work, we present the analysis results of charmonium properties at both the correlator and spectral function level. Previous work has been reported in Ref.~\cite{Ding09_1,Ding09_2}.


 The charmonium correlators at vanishing momentum are calculated using:
 \be
G_{H}(\tau,T)=\sum_{\vec{x}} \langle~ J_H(\tau,\vec{x})~J_{H}^{\dag}(0,\vec{0})~\rangle_T .
\ee
 $J_H$ is a suitable mesonic operator, here we consider a local operator of $\bar{q}(\tau,\vec{x})\Gamma_{H} q(\tau,\vec{x})$, where $\Gamma_{H}=\gamma_{i},\gamma_{5}$ for vector ($V_{ii}$) and pseudo-scalar ($PS$) channels, respectively. The temperature $T$ is related to Euclidean temporal extent $aN_{\tau}$ by $T=1/(aN_{\tau})$, where $a$ is the lattice spacing. Through analytic calculation, the correlation function can be related to the spectral function as the following:
\be
G_{H}(\tau,T)=\int_0^{\infty}{\mathrm{d}\omega~\sigma_{H}(\omega,T)}~K(\tau,T,\omega),
\label{eq:relation_cor_spf}
\ee
where the kernel $K$ is given by $K(\tau,T,\omega)= \mathrm{cosh}(\omega(\tau-\frac{1}{2T}))/\mathrm{sinh}(\frac{\omega}{2T})$.
The spectral function $\sigma(\omega)$ contains all the information of the hadron properties in the medium and is the key quantity to be investigated.
For instance, the dissociation temperature can be read from the deformation of the spectral function and the heavy quark diffusion constant $D$ relates to the vector spectral function as
\be
D= \frac{\pi}{3\chi_{00}}\lim_{\omega\rightarrow0}\sum_{i=1}^{3}\frac{\sigma_{V}^{ii}(\omega,T)}{\omega},
\label{HQ_diff}
\ee
where $\chi_{00}$ is the quark number susceptibility.

Inverting Eq.~(\ref{eq:relation_cor_spf}) to obtain the spectral function at finite temperature is hampered mainly by two issues: the temporal extent is always restricted by the temperature, $a\tau\leq1/T$;  the spectral functions we want to have should be continuous and have a degree of freedom of $\mathcal{O}$(1000) but the correlators are calculated in the discretized time slices with limited numbers, typically $\mathcal{O}$(10), which makes the inversion  ill-posed. Thus the normal $\chi^2$ fitting would be inconclusive. Maximum Entropy Method (MEM) is currently one of the best tools in the literature to solve the problem~\cite{Asakawa01}. It is based on the Bayesian algorithm and requires the prior knowledge of spectral functions as an input. The only parameter in the MEM analysis is the so called ``default model", which provides the prior information of the spectral function, e.g. the spectral function should be positive-definite.


 \begin{table}[htdp]
\begin{center}
\newcolumntype{R}{>{\raggedleft\arraybackslash}X}
\newcolumntype{L}{>{\raggedright\arraybackslash}X}
\begin{tabular}{ccccccccc}
\hline
\hline
$\beta$       & $a$ [fm]  &   $a^{-1}$[GeV]     &  $L_{\sigma}$ [fm]&$c_{\rm SW}$   &   $\kappa$   &      $N_{\sigma}^{3} \times N_\tau$  &      T/$\Tc$    & $N_{conf}$    \\
\hline
 7.793           &  0.010     & 18.974                & 1.33                             & 1.310381    &    0.13200    &         $128^{3} \times 96 $                &      0.73      &   234   \\
                      &                  &                              &                                      &                      &                      &          $128^{3} \times 48 $                &      1.46        &     461 \\
                       &                 &                              &                                     &                      &                      &           $128^{3} \times 32 $                &      2.20      &  105\\
                        &                &                              &                                      &                      &                      &          $128^{3} \times 24 $                &      2.93       & 81 \\
                        \hline
\hline
\end{tabular}
\end{center}
\caption{Lattice parameters.}
\label{table:lattice_parameters}
\end{table}

The standard Wilson plaquette action for the gauge field and the non-perturbatively $\cO(a)$ improved clover fermion action for charm quarks are implemented in the simulation. 
 The mass of vector meson is tuned to the physical $\Jpsi$ mass.  We measured correlation functions on very fine quenched lattices with a relatively large size of $128^{3}\times 96$, $128^3\times48$, $128^3\times32$ and $128^3\times24$ at $0.73~T_c$, $1.46~T_c$, $2.20~T_c$ and $2.93~T_c$, respectively. The lattice parameters are shown in Table~\ref{table:lattice_parameters}.

We first analyze the temperature dependence of charmonium states at the correlator level. One constructs~\cite{Datta04}
\be
G_{\rec}(\tau,T;T^\prime) = \int_0^{\infty}\md \omega~\sigma(\omega,T^\prime)\,\,\frac{\cosh\left(\omega(\tau - 1/2T)\right)}{\sinh(\omega/2T)},
\label{rec_cor}
\ee
to study the difference of the spectral function at temperature $T$ and $T^{\prime}$. The deviation of $G(\tau,T)$ from $G_{\rec}(\tau,T;T^\prime)$ indicates any modifications of the spectral function at temperatures $T$ from the one at temperature $T^\prime$. One normally needs  a technique to obtain the spectral function $\sigma(\omega,T^{\prime})$ at a reference temperature $T^{\prime}$ and consequently the evaluation of
Eq.~(\ref{rec_cor}) suffers from the uncertainty of the determination of the spectral function brought by the certain technique. We found a useful exact relation, which is a generalization of Ref.~\cite{Meyer10}, as follows:
\be
 \frac{\cosh[\omega(\tau-N_{\tau}/2)]}{\sinh (\omega N_{\tau}/2)}~~\equiv~\sum_{\tau^{\prime}=\tau;~\tau^{\prime}+=N_{\tau}}^{N_{\tau}^{\prime}-N_{\tau}+\tau} \frac{\cosh[\omega(\tau^{\prime}-N_{\tau}^{\prime}/2)]}{\sinh (\omega N_{\tau}^{\prime}/2)},
\label{kernel_rules}
\ee
where $T^{\prime}=(a N_{\tau}^{\prime})^{-1},~~T=(aN_{\tau})^{-1},~~\tau^{\prime}\in[0,~N_{\tau}^{\prime}-1],~~\tau\in[0,~N_{\tau}-1],~~N_{\tau}^{\prime}=m~ N_{\tau},~~m\in\mathbb{Z}^{+}$. $N_{\tau}$ and $N_{\tau}^{\prime}$ are the number of time slices in the temporal directions at temperature $T$ and $T^{\prime}$, respectively. $\tau$ denotes the time slice of the correlation function at temperature $T$ while $\tau^{\prime}$ denotes the time slice of the correlation function at 
temperature $T^{\prime}$. The sum of $\tau^{\prime}$ on the right hand side of  Eq.~(\ref{kernel_rules}) starts from $\tau^{\prime}=\tau$ with a step length $N_{\tau}$ to the upper limit $N_{\tau}^{\prime}-N_{\tau}+\tau$. After putting $\sigma(\omega,T^{\prime})$ into both sides of the above relation and performing the integration over $\omega$, one immediately arrives at:

\be
G_{\rec}(\tau,T;T^{\prime}) = \sum_{\tau^{\prime}=\tau;~\tau^{\prime}+=N_{\tau}}^{N_{\tau}^{\prime}-N_{\tau}+\tau}  G(\tau^{\prime},T^{\prime}),
\label{eq:Grec_data}
\ee
which shows the evaluation of $G_{\rec}(\tau,T; T^{\prime})$ can be done directly from the correlator $G(\tau^{\prime},T^{\prime})$ at $T^{\prime}$. In what follows, we suppress
$T^{\prime}$ in $G_{\rec}(\tau,T; T^{\prime})$.
 \begin{figure}[htbl]
 \centering
 \includegraphics[width=.5\textwidth]{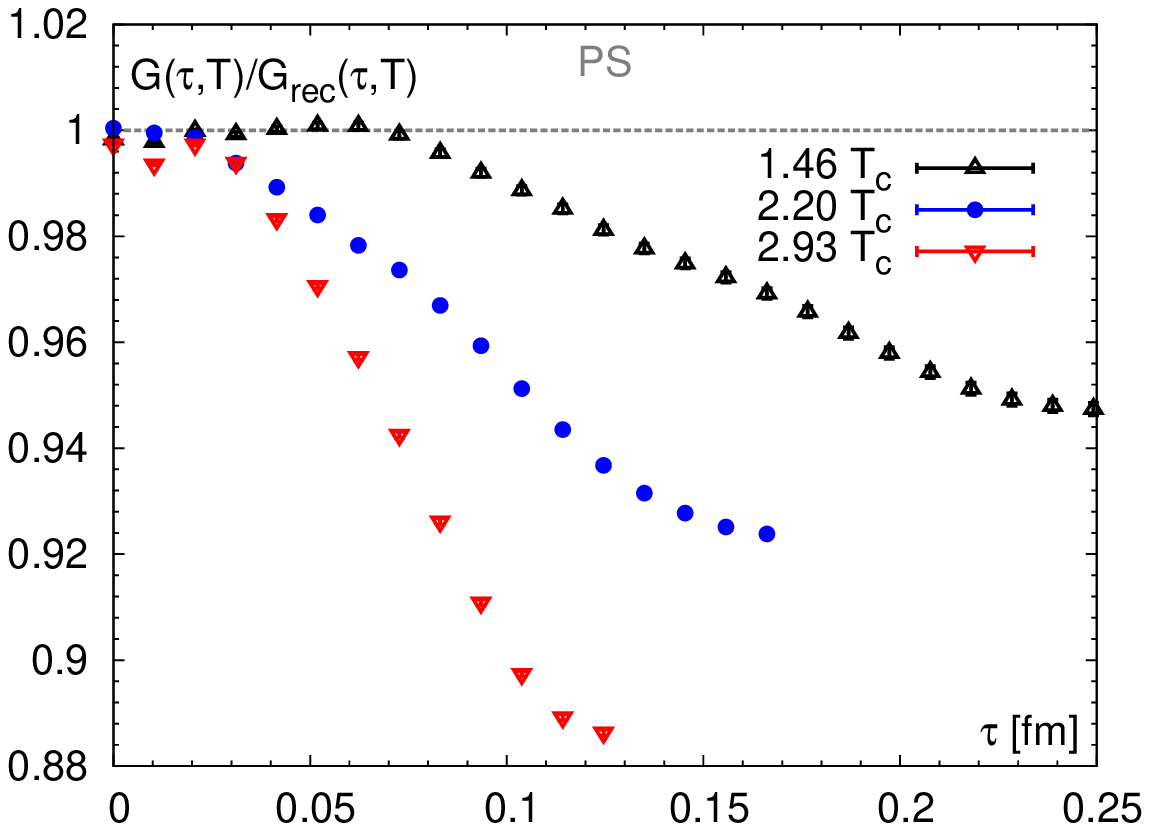}~\includegraphics[width=.5\textwidth]{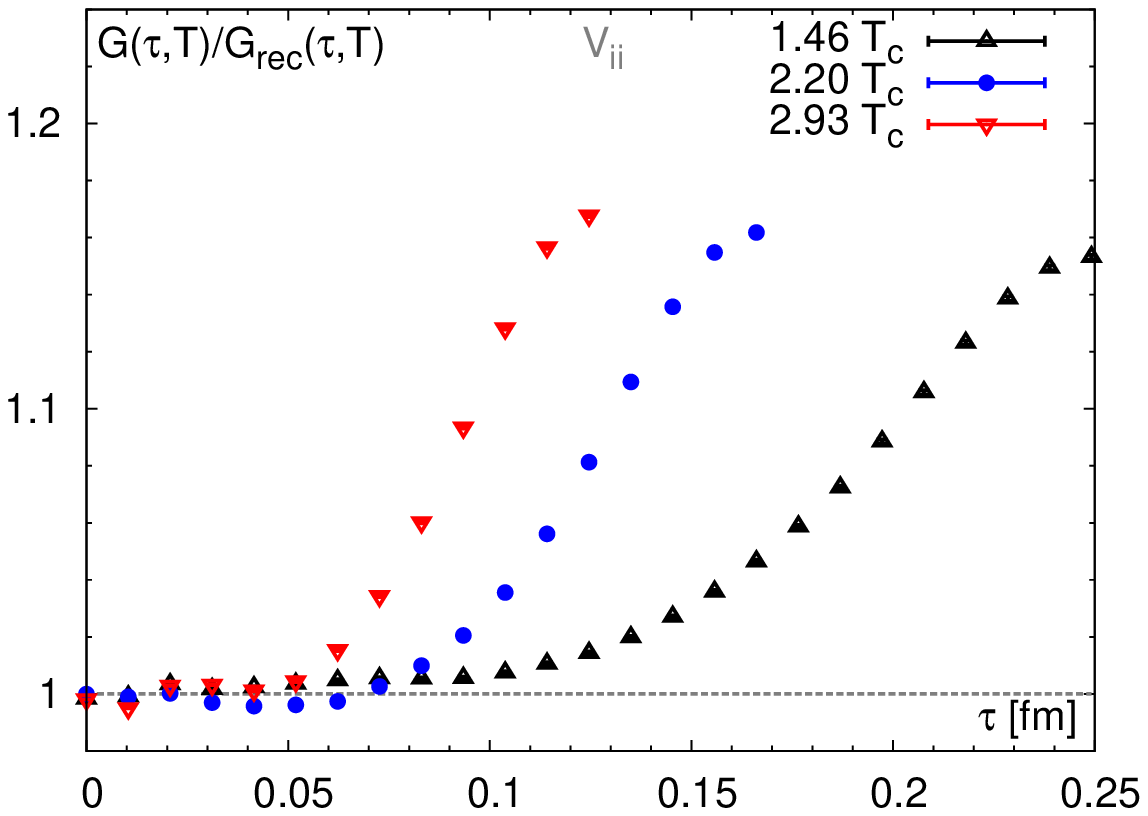}~
\caption{The ratio $G(\tau,T)/G_{\rm rec}(\tau,T)$ in $PS$ (left) and $V_{ii}$ (right) channels at $T=1.46$, 2.20 and 2.93$~T_c$.}
 \label{fig:GoverGrec_Swave_std_beta7p793}
\end{figure}

The ratios $G(\tau,T)/G_{\rm rec}(\tau,T)$ are shown in Fig.~\ref{fig:GoverGrec_Swave_std_beta7p793}. The $G_{\rec}$ are obtained from the correlator data at $0.73~T_c$ through Eq.~(\ref{eq:Grec_data}). In the left plot of Fig.~\ref{fig:GoverGrec_Swave_std_beta7p793} one can see that the ratios in the $PS$ channel at all temperatures decrease monotonically with increasing distance. The temperature effects set in at a shorter distance at a higher temperature. At the largest distances, the ratios deviate from unity at around 5\%, 8\% and 12\% at 1.46, 2.20 and 2.93 $T_c$, respectively. The results for the $V_{ii}$ channel are shown in the right plot of Fig.~\ref{fig:GoverGrec_Swave_std_beta7p793}. Different from the case in the $PS$ channel, the ratios increase monotonically with increasing distance. At the largest distances, the ratios have a larger deviation from unity compared with the results in the $PS$ channel. The deviation at the largest distance at $1.46~T_c$ is around 16\% and it is comparable to the values at higher temperatures.
 \begin{figure}[htbl]
 \centering
 \includegraphics[width=.5\textwidth]{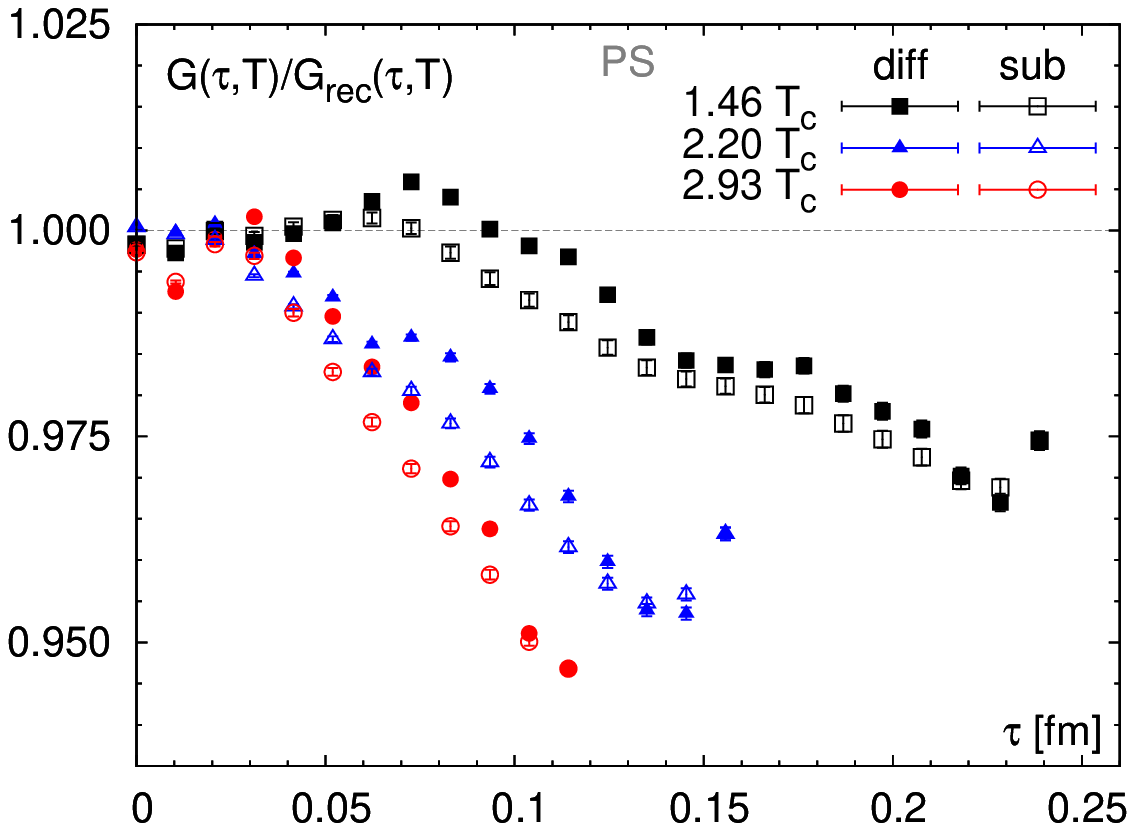}~\includegraphics[width=.5\textwidth]{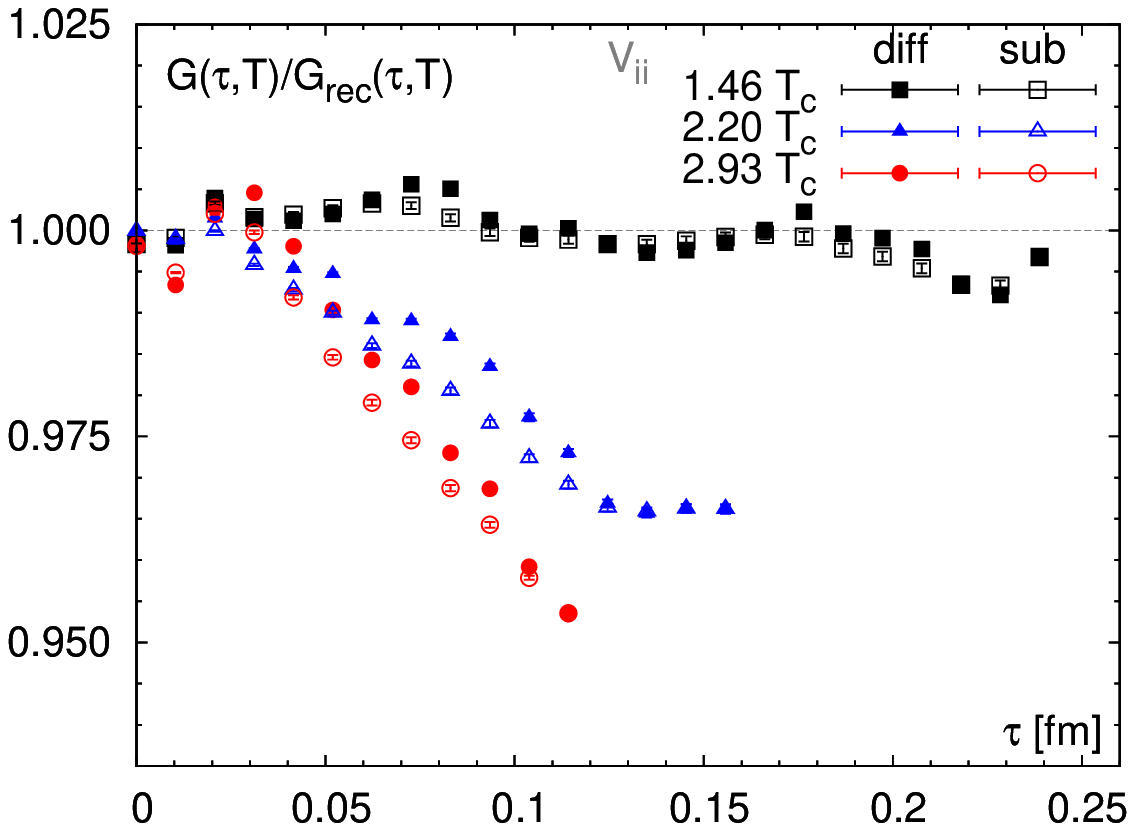}~
\caption{The ratio $G^{\rm sub}(\tau,T)/G^{\rm sub}_{\rm rec}(\tau,T)$ ($G^{\rm diff}(\tau,T)/G^{\rm diff}_{\rm rec}(\tau,T)$) in $PS$ (left) and $V_{ii}$ (right) channels at $T=1.46$, 2.20 and 2.93$~T_c$. The ``diff" and ``sub" stand for the results of the ratio $G^{\rm diff}/G^{\rm diff}_{\rm rec}$ and $G^{\rm sub}/G^{\rm sub}_{\rm rec}$. }
 \label{fig:GoverGrec_Swave_diff_sub_beta7p793}
\end{figure}

The large temperature dependence of $G(\tau,T)/G_{\rec}(\tau,T)$ in the $V_{ii}$ channel could be due to the possible zero mode contributions. To suppress the zero mode contribution we evaluate the ratio of the differences of the neighboring correlators to the difference of the corresponding reconstructed correlators~\cite{Petreczky08}

\be
\frac{G^{\rm diff}(\tau,T)}{G^{\rm diff}_{\rm rec}(\tau,T)} \equiv \frac{G(\tau,T) -  G(\tau+1,T)}{G_{\rm rec}(\tau,T) -  G_{\rm rec}(\tau+1,T)},
\label{eq:GoverGrec_diff}
\ee
which equals the ratio of the time derivative of the correlators to the time derivative of the reconstructed correlators at $\tau+1/2$. One can also check the ratio of midpoint subtracted correlators~\cite{Umeda07}
\be
\frac{G^{\rm sub}(\tau,T)}{G^{\rm sub}_{\rm rec}(\tau,T)} \equiv \frac{G(\tau,T) -  G(N_{\tau}/2,T)}{G_{\rm rec}(\tau,T) -  G_{\rm rec}(N_{\tau}/2,T)}.
\label{eq:GoverGrec_midpoint-subtracted}
\ee
In those two ratios a $\tau$ independent constant cancels in the correlator.

The results for~$\RGsub $ and $\RGdiff $ in $PS$ (left) and $V_{ii}$ (right) channels are shown in Fig.~\ref{fig:GoverGrec_Swave_diff_sub_beta7p793}. The ratios $\RGsub$ and $\RGdiff$ give similar results at all the distances. Seen from the right panel of Fig.~\ref{fig:GoverGrec_Swave_diff_sub_beta7p793} the magnitude of the measured vector correlator to the reconstructed correlator reduces dramatically after the implementation of the difference of neighboring correlators (Eq.~(\ref{eq:GoverGrec_diff})) and mid-point subtracted correlators (Eq.~(\ref{eq:GoverGrec_midpoint-subtracted})). At $1.46~T_c$ the ratio is more or less unity at all distances, at $2.20~T_c$ and $2.93~T_c$ the ratio becomes even smaller than unity at large distances. The zero mode contribution accounts for most temperature dependence of the ratio $G/G_{\rm rec}$ at least at $1.46~T_c$ seen in Fig.~\ref{fig:GoverGrec_Swave_std_beta7p793}. The deviations of the ratios from unity in the PS channel shown in the right panel of Fig.~\ref{fig:GoverGrec_Swave_diff_sub_beta7p793} are also reduced. However, the effect is not as strong as that in the $V_{ii}$ channel and the values at the largest distance are shifted up only about 3\% at both $1.46~T_c$ and $2.20~T_c$ and about 6\% at $2.93~T_c$ compared with the results in Fig.~\ref{fig:GoverGrec_Swave_std_beta7p793}. Comparing the results for the $V_{ii}$ channel with those for the $PS$ one in Fig.~\ref{fig:GoverGrec_Swave_diff_sub_beta7p793}, we find the ratios $\RGsub$ ($\RGdiff$) in these two channels have similar behavior at all distances at two higher temperatures $2.20~T_c$ and $2.93~T_c$. However, they differ at $1.46~T_c$. The phenomenon we observe here might suggest that $\Jpsi$ could survive up to $1.46~T_c$ and starts to melt at $2.20~T_c$, and $\eta_c$ might be melted already at $1.46~T_c$.

One has to note that the comparison of the measured correlator with the reconstructed correlator can only give a rough idea of the magnitude of any medium 
effects at a certain temperature. To really explore the properties of the charmonium states, one has to go to the spectral function level. Thus it is crucial to extract the spectral function from the correlators using MEM.

\begin{figure}[!t]
  \begin{center}
    \includegraphics[width=0.5\textwidth]{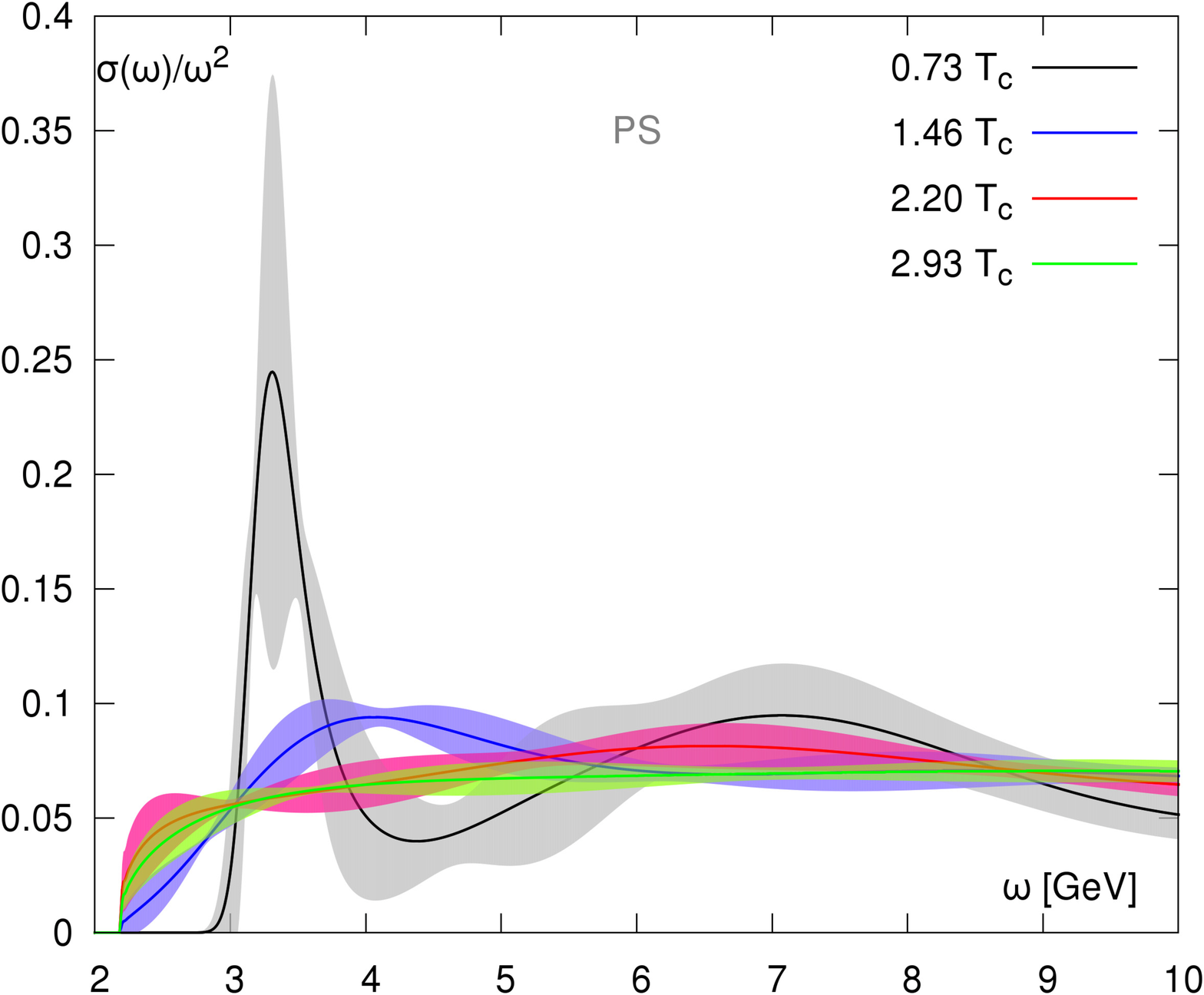}~\includegraphics[width=0.5\textwidth]{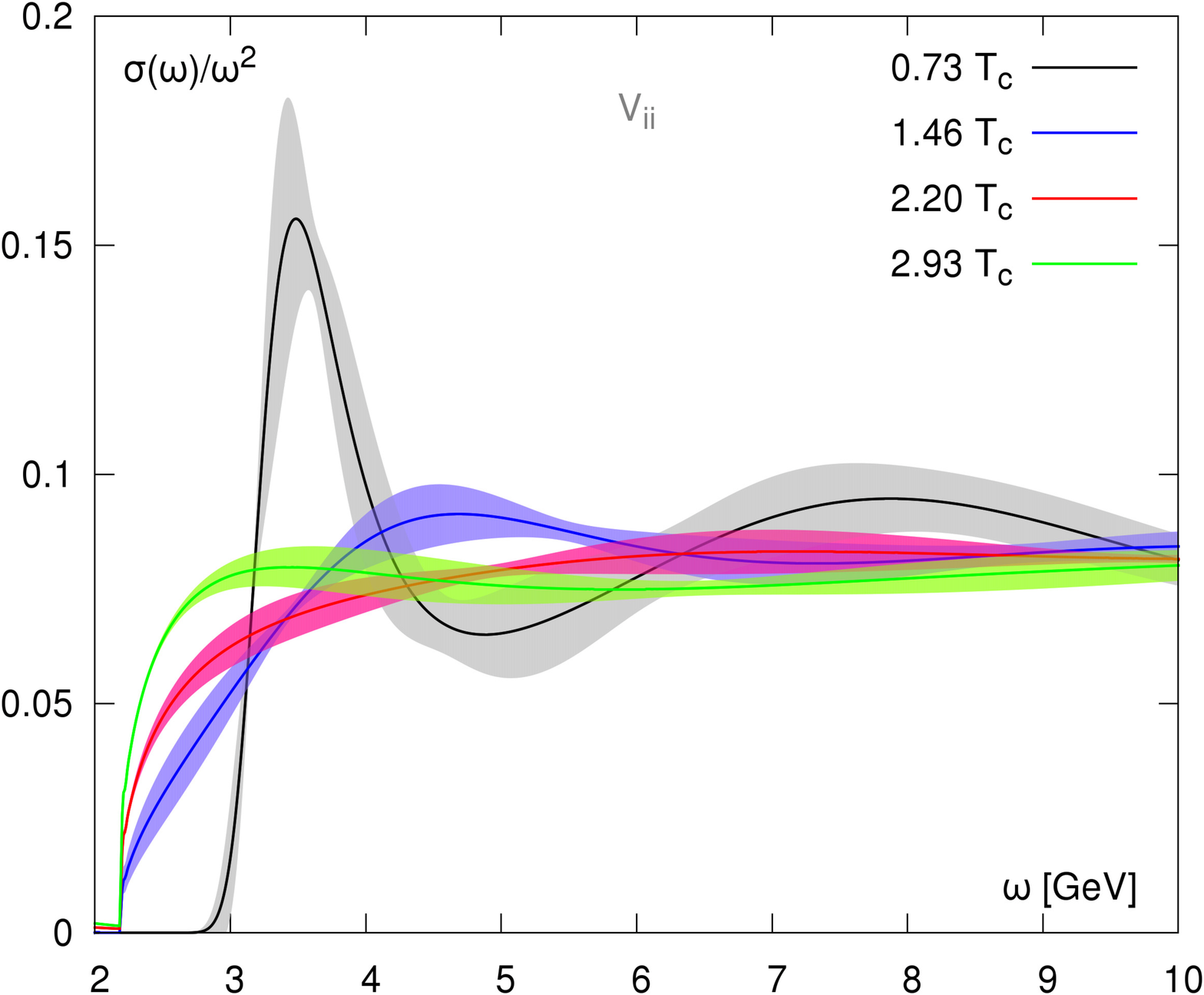}~
                    \caption{The statistical errors for the output spectral functions in $PS$  (left) and $V_{ii}$ (right) channels at all available temperatures. 
                    The shaded areas are the errors of the output spectral functions. The mean values are the solid lines inside the shaded areas.
          }
                 \label{fig:spf_ps_statistics_errors}
  \end{center}
\end{figure}
\begin{figure}[!t]
  \begin{center}
  \includegraphics[width=0.5\textwidth]{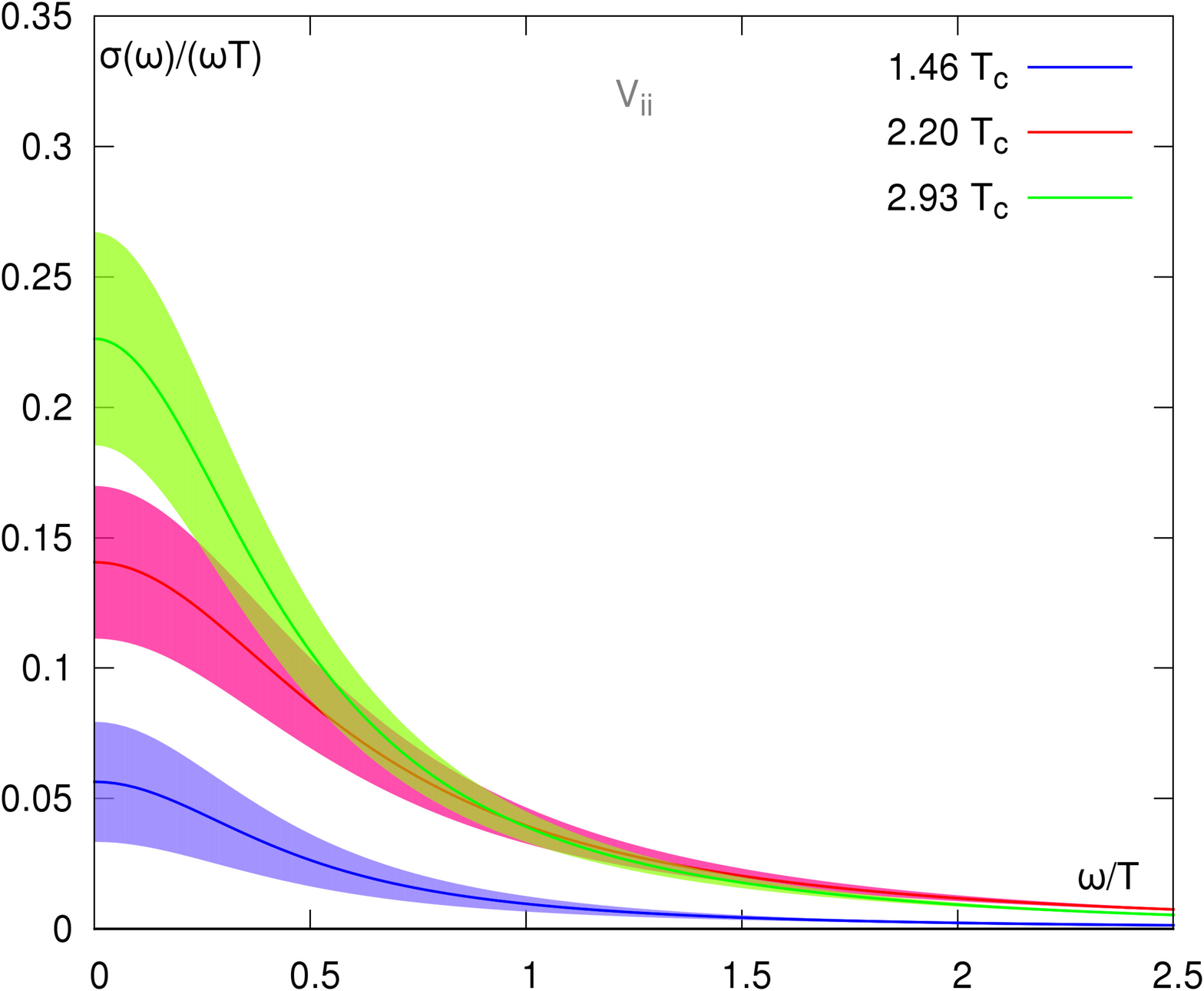}~ \includegraphics[width=0.5\textwidth,height=0.43\textwidth]{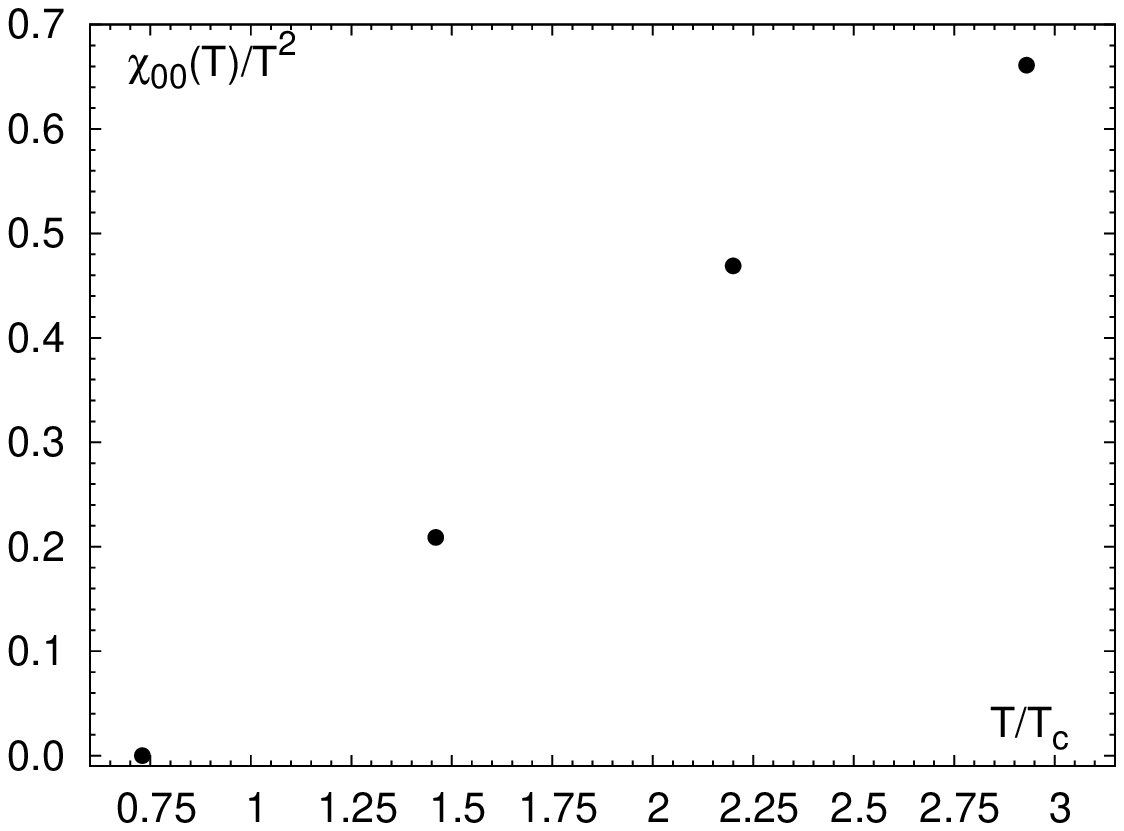}~
                \caption{Left: the statistical errors for the transport part of output spectral functions in the $V_{ii}$ channel at all available temperatures. The shaded areas are the errors of the output spectral function. The mean values are the solid lines inside the shaded areas. Right: the quark number susceptibility $\chi_{00}/T^2$ versus $T/T_c$.
}
                 \label{fig:spf_trans}
\end{center}
\end{figure}

 In the MEM analysis, we use number of points in the investigated energy region $N_{\omega}=8000$, the minimum energy $a\omega_{min}=0.000001$ and implement the modified kernel $K^{\ast}=\mathrm{tanh}(\omega/2)\cdot K$~\cite{Ding09_2,Engels10} to explore the low energy behavior of spectral function~\cite{Aarts07}. The default model we used in the $PS$ channel is a normalized free lattice spectral function and in the $V_{ii}$ channel is a normalized free lattice spectral function with a transport peak modeled by a Breit-Wigner distribution at small $\omega$. Here we show the spectral functions in $PS$ and $V_{ii}$ channels with statistical uncertainties\footnote{The systematic error analyses have been performed in Ref.~\cite{Ding_Dissertation}, which do not change the general results shown in Fig.~\ref{fig:spf_ps_statistics_errors} and Fig.~\ref{fig:spf_trans}.}.

The statistical error is obtained from the Jackknife method. Quite often the statistical error is given on the integral of the spectral function $\sigma(\omega)$ over a certain $\omega$ region in the spectral function plot~\cite{Datta04,lattice_app}. However, it is not straightforward to get a feeling of how big the error is
on the spectral function itself. Here we rather calculate the Jackknife error on each point of the spectral function. We show the results for $PS$ (left) and $V_{ii}$ (right) channels in the intermediate frequency region in Fig.~\ref{fig:spf_ps_statistics_errors} and show the transport part of spectral functions in the $V_{ii}$ channel in Fig.~\ref{fig:spf_trans}. In both figures, shaded areas denote the Jackknife errors  and solid lines inside the shaded areas are the mean values of spectral functions.

From the left plot in Fig.~\ref{fig:spf_ps_statistics_errors} one can see that at $0.73~T_c$ the spectral function in the $PS$ channel has large uncertainties in the amplitude at the point which corresponds to the ground state peak location in the mean spectral function. However, even at the lower end of the error bar, the amplitude is still larger than the peak amplitudes at the higher temperatures within the errors. The peak location of the ground state peak at $0.73~T_c$ might be shifted to a lower energy of $\omega\approx3~$GeV or to a higher energy at $\omega\approx3.6~$GeV. In the latter case, the peak location would have the same peak location as the spectral function at $1.46~T_c$ but with a much larger amplitude and smaller width. At $2.23~T_c$ there is hardly a peak structure within the statistical errors. At $2.93~T_c$ the spectral function flattens. Thus this picture suggests $\eta_c$ is ``partly" melted at $1.46~T_c$ and dissolves at higher temperatures. In the right plot of Fig.~\ref{fig:spf_ps_statistics_errors}, we focus on the resonance part of the spectral function in the $V_{ii}$ channel. One sees that the peak location of the spectral function at $1.46~T_c$ does not have an overlap with the peak location of the spectral function at $0.73~T_c$. The amplitudes between these two differ a lot. At both $2.20~T_c$ and $2.93~T_c$ there are hardly any peak structures and at $2.93~T_c$ the spectral function is flattened. This picture indicates $\Jpsi$ is already melted at $1.46~T_c$. 

The statistical uncertainties of the transport peaks in the $V_{ii}$ channel are shown in the left plot of Fig.~\ref{fig:spf_trans}. The amplitude of the transport peak at $\omega=0$ gives the value of the heavy quark diffusion constant. The uncertainties of both, amplitudes and widths of the peak, are relatively small.  Through Eq.~(\ref{HQ_diff}) and $\chi_{00}/T^2$ shown in the right plot of Fig.~\ref{fig:spf_trans}, we get $DT$ very roughly to be $0.28$ at 1.46 $T_c$ and find $DT$ increases with increasing temperatures. The precise determination of $DT$ needs further detailed study.

In summary, our analysis suggests that $\Jpsi$ is melted already at $1.46~T_c$ and $\eta_c$ starts to dissolve at $1.46~T_c$ and does not exist at higher temperatures. We also identify the transport contribution at the spectral function level for the first time. 


\begin{acknowledgments}
This work has been supported in part by the Deutsche Forschungsgemeinschaft under grant GRK 881 and by contract DE-AC02-98CH10886 with the U.S. Department of Energy. Numerical simulations have been performed on the BlueGene/P at the New York Center for Computational Sciences (NYCCS) which is supported by the State of New York and the BlueGene/P at the John von Neumann Supercomputer Center (NIC) at FZ-J\"ulich, Germany.

\end{acknowledgments}


\end{document}